\documentclass{article}
\usepackage[T1]{fontenc} % add special characters (e.g., umlaute)
\usepackage[utf8]{inputenc} % set utf-8 as default input encoding
\usepackage{ismir,amsmath,cite,url}
\usepackage{graphicx}
\usepackage{color}

\usepackage{comment}
\usepackage{subcaption}
\usepackage{tabularx}
\usepackage{float}
\usepackage{adjustbox}

\usepackage{lineno}

\usepackage{float}
\usepackage{siunitx}

\usepackage{lineno}
\title{Deep Learning Based Source Separation Applied To Choir Ensembles}

\multauthor{Darius Petermann$^1$ \hspace{1cm} Pritish Chandna$^1$ \hspace{1cm} Helena Cuesta$^1$ } {\bfseries{Jordi Bonada$^1$ \hspace{1cm} Emilia G\'omez$^{2,1}$}\\ $^1$ Music Technology Group, Universitat Pompeu Fabra, Barcelona\\
$^2$ European Commission, Joint Research Centre, Seville\\
\tt\small dariusarthur.petermann01@estudiant.upf.edu,\\ \tt\small \{pritish.chandna, helena.cuesta, jordi.bonada, emilia.gomez\}@upf.edu
}

\def\authorname{D. Petermann, P. Chandna, H. Cuesta, J. Bonada, and E. G\'omez}

\begin{document}

\maketitle
\begin{abstract}

Choral singing is a widely practiced form of ensemble singing wherein a group of people sing simultaneously in polyphonic harmony. The most commonly practiced setting for choir ensembles consists of four parts; Soprano, Alto, Tenor and Bass (SATB), each with its own range of fundamental frequencies (F$0$s). The task of source separation for this choral setting entails separating the SATB mixture into the constituent parts. Source separation for musical mixtures is well studied and many deep learning based methodologies have been proposed for the same. However, most of the research has been focused on a typical case which consists in separating vocal, percussion and bass sources from a mixture, each of which has a distinct spectral structure. In contrast, the simultaneous and harmonic nature of ensemble singing leads to high structural similarity and overlap between the spectral components of the sources in a choral mixture, making source separation for choirs a harder task than the typical case. This, along with the lack of an appropriate consolidated dataset has led to a dearth of research in the field so far. In this paper we first assess how well some of the recently developed methodologies for musical source separation perform for the case of SATB choirs. We then propose a novel domain-specific adaptation for conditioning the recently proposed U-Net architecture for musical source separation using the fundamental frequency contour of each of the singing groups and demonstrate that our proposed approach surpasses results from domain-agnostic architectures.

\end{abstract}
\section{Introduction}
\label{sec:introduction}

Choir music is a well-established and long-standing practice involving a body of singers performing together. Such ensembles are usually referred to as choir and may perform with or without instrumental accompaniment. A choir ensemble is usually structured by grouping the voices into four different sections, each depicting different frequency ranges for the singers; "Soprano" (\SI{260}{\hertz}-\SI{880}{\hertz}), "Alto" (\SI{190}{\hertz}-\SI{660}{\hertz}), "Tenor" (\SI{145}{\hertz}-\SI{440}{\hertz}), and "Bass" (\SI{90}{\hertz}-\SI{290}{\hertz})~\cite{SATB_RANGE}. This type of structural setting is usually referred to as a SATB setting. Although different variants of this structure exist, the SATB is the most well documented, with several conservatories across Europe dedicated to the study and practice of the art form, highlighting its cultural significance. This will be the main focal point of our study.

The segregation of a mixture signal into its components is a well researched branch of signal processing, known as source separation. For polyphonic music recordings, this implies the isolation of the various instruments mixed together to form the whole. With applications such as music remixing, rearrangement, audio restoration, and full source extraction, its potential use in music is of great appeal. While the task remains similar independently of the type of setting involved, the nature of the sources (e.g.: speech, musical instrument, singing voice) and their relations may entail various challenges and, consequently, require different separation methodologies to be employed.

The most studied case of musical source separation focuses on pop/rock songs, which typically have three common sources; vocals, drums, bass along with other instrumental sources which are usually grouped together as others. A large body of research~\cite{MONO_SIGSEP2017,UMIX2019,ENTOEND2018} has been published in this field over the last few years, beginning with the consolidation of a common dataset for researchers to train and evaluate their models on. In 2016, \textit{DSD100}~\cite{SiSEC16} was first introduced and made available to the public and was later extended to \textit{MUSDB18}~\cite{MUSDB18}, which comprises \num{150} full-length music tracks for a total of approximately \num{10} hours of music. To this day, \textit{MUSDB18} represents the largest freely available dataset of its kind. 

While source separation for the pop/rock case has come leaps and bounds in the last few years, it remains largely unexplored for the SATB choir case, despite its cultural importance. This is partly due to the lack of a consolidated dataset, similar to the \textit{MUSDB18}, and partly due to the nature of the task itself. The sources to be separated in pop/rock have distinct spectral structure; the voice is a harmonic instrument and has a distinct spectral shape, defined by a fundamental frequency and its harmonic partials and formants. The bass element to be separated also has a harmonic structure, but lacks the formants found in the human voice and has a much lower fundamental frequency than the human voice. In contrast, the spectrum of a percussive instrument is generally inharmnoic and energy is usually spread across the spectrum. In contrast, the sources to be separated in a SATB choir all have a similar spectral structure with a fundamental frequency, partials and formants. This makes the task more challenging than its more studied counter part. However, the distinct ranges of fundamental frequencies in the sources to be separated can be used to distinguish between them, a key aspect that we aim to explore in our study. 

We build on top of some recently proposed Deep Neural Network (DNN) models to separate SATB monoaural recordings into each of their respective singing groups and then propose a specific adaptation to one of the models. The rest of the paper is organized as follow: Section~\ref{sec:sota} presents and investigates some of the recently proposed high performance deep learning based algorithms used for common musical source separation tasks, such as the U-Net~\cite{UNET2015} architecture and its waveform-based adaptation, Wave-U-Net~\cite{WAVEUNET2018}. Section~\ref{sec:datset} goes over the dataset curation carried out for this experiment. Section~\ref{sec:method} presents our adaptation of the conditioned U-Net model described in~\cite{CUNET2019}, with a control mechanism conditioned on the input sources' fundamental frequency (F$0$). Section~\ref{sec:eval_methodology} defines the evaluation metrics and methodology used in this experiment. In Section~\ref{sec:results} we evaluate and compare how existing models and our proposed adaptation perform on the task of source separation for SATB recordings. We then present and discuss the results. Section~\ref{sec:conclusion} finally concludes with a discussion around our experiment and provide comments on future research that we intend to carry out.

\section{Related Work}
While source separation has remained relatively unexplored for the case of SATB choirs, a number of architectures have been proposed over the last few years for musical source separation in the pop/rock case. A comprehensive overview of all proposed models is beyond the scope of this study, but we provide a summary of some of the most pertinent models that we believe can easily be adapted to the case in study. 
\label{sec:sota}
\subsection{U-Net}

The U-Net architecture~\cite{UNET2015}, which was specifically developed to process and segment biomedical images, inspired many subsequent audio-related adaptations due to its unprecedented performance.

The original model includes an encoding path, which reduces the initial input into a latent representation (bottleneck) followed by a decoding path, which expends the channels’ receptive field back into its original shape while concatenating the feature maps from the contracting path by the mean of skip connection layers.

One of the first paper to present a U-Net adaptation towards audio source separation was proposed by Jansson et al.~\cite{SING_UNET2017}, where they propose an architecture which specifically targets vocal separation performed on western commercial music (or pop music). The authors present an architecture directly derived from the original U-Net one, which takes spectrogram representations of the sources as input and aims at predicting a soft-mask for the targeted source (either vocal or instrumental). The predicted mask is then multiplied element-wise with the original mixture spectrogram in order to obtain the predicted isolated source. It is worth mentioning that for each of the given sources, a U-Net instance is trained in order to predict its respective mask. In the case of SATB mixtures, four U-Net instances are necessary in order to predict each of the four singing groups.

\subsection{Conditioned-U-Net}

Depending on the nature of the separation task, its underlying process can easily lead to scaling issues. The conditioned U-Net (C-U-Net) architecture, described in~\cite{CUNET2019}, aims at addressing this limitations by introducing a mechanism controlled by external data which govern a single U-Net instance. C-U-Net does not diverge much from the initial U-Net one; as an alternative to the multiple instances of the model, each of which is specialized in isolating a specific source, C-U-Net proposes the insertion of feature-wise linear modulation (\textit{FiLM}) layers~\cite{FILM2017}, which represents an affine transform defined by two scalars - \(\gamma{}\) and \(\beta{}\), across the architecture. This allows for the application of linear transformations to intermediate feature maps. These specialized layers conserve the shape of the original intermediate feature input while modifying the underlying mapping of the filters themselves. 

\begin{equation}
    \label{film_layer}
     \displaystyle FiLM(x) = \gamma{(z)} \dot{x + \beta{(z)}}\
\end{equation}

\begin{figure}
    \centering
    \includegraphics[width=0.3\textwidth]{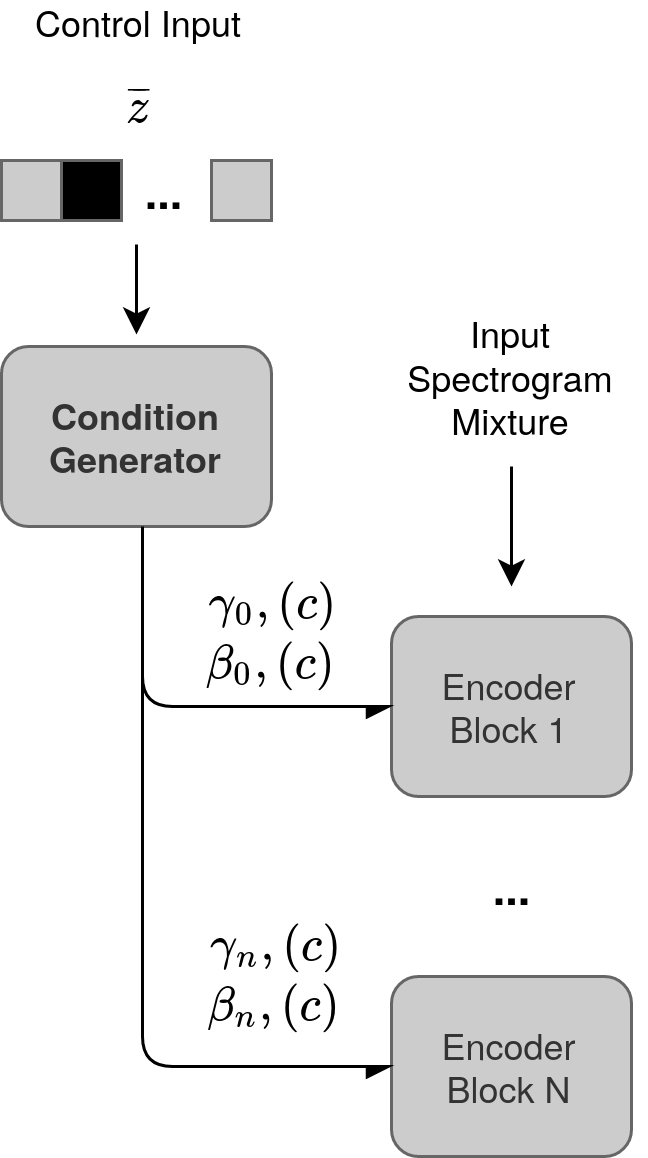}
    \caption{C-U-Net control mechanism with the vector $\protect\overline{z}$, a one-hot representation of the source to separate, which dictates the N sets of $\protect\gamma$ and $\protect\beta$ values to be used by the \textit{FiLM} layer at each of the block of the encoding path, \num{1} to N.}
  \label{fig:cunet_mechanism}
\end{figure}

In eq. \eqref{film_layer}, \(x\) is the input of the \textit{FiLM} layer, \(\gamma\) and \(\beta\) the parameters that scale and shift \(x\) based on an external information, \(\overline{z}\)~\cite{CUNET2019}. \(\gamma{}\) and \(\beta{}\) modulates the feature maps according to an input vector \(\overline{z}\), which describes the source to separate. The condition generator block described in Figure \ref{fig:cunet_mechanism} represents a neural network embedding the one-hot encoding input \(\overline{z}\) into the most optimal values to be used by the \textit{FiLM} layer.

\subsection{Wave-U-Net}

In~\cite{WAVEUNET2018}, the authors present a time-domain adaptation of the U-Net architecture, which performs the separation operation on the waveform. As the input is a one-dimensional signal, the feature maps are computed directly from the waveform samples through \num{1}D convolution operations. Because Wave-U-Net takes raw waveforms as input, the initial U-Net model has to be adapted accordingly in order to accommodate for the input's nature. Consequently, the feature maps along both contracting and expanding paths are computed by the means of single-dimensional convolution layers. Both paths contain twelve convolutional layers, each, for a total of \num{24} layers. In the down-sampling path, the receptive field is reduced in half after each layer while the input feature maps are increased by a factor of \num{24} every time. On the other hand, in the up-sampling path the time-context is doubled after every convolutional layer while the feature maps are reduced, again by \num{24}, after every layers. By that mean, the receptive field and the number of channels of the original input signal will remain preserved at the output stage. Although Wave-U-Net has proven to deliver satisfying results on common musical source separation tasks, the fact remains that waveform-based architectures in general require more data than their spectrogram-based counter parts.

\section{DATASET}
\label{sec:datset}
The training data we have curated for this experiment are composed of the following two datasets:

\begin{itemize}
    \item Choral Singing Dataset~\cite{CUESTA_DST2018} (CSD). Three songs performed by \num{16} singers from the Anton Bruckner Choir (SATB)\footnote{\url{https://zenodo.org/record/1286570#.XyGcHy-z3yU}}.
    \item A proprietary dataset with \num{26} Spanish SATB songs by \num{4} singers, one for each part.
\end{itemize}

There are very few publicly available choir music datasets, thus our choice remains limited. To this day and to the best of our knowledge, there isn't any existing dataset which is specifically suited to our task, thus one of the subsidiary work of this experiment revolves around curating a proper and complete dataset to train our various models. For our experiment, we take advantage of the CSD~\cite{CUESTA_DST2018}. This dataset was recorded in a professional studio and contains individual tracks for each of the \num{16} singers of a SATB choir, i.e. \num{4} singers per choir section. It comprises three different choral pieces: \textit{Locus Iste}, written by Anton Bruckner, \textit{Nīno Dios d’Amor Herido}, written by Francisco Guerrero, and \textit{El Rossinyol}, a Catalan popular song; all of them were written for \num{4}-part mixed choir. The dataset is very well suited for our experiment as the isolated track for each individual singer will allow us to proceed the same way as in~\cite{MULTIF0_SATB2019}, that is to create artificial mixes by combining various stems from different groups together. Using different combinations of all \num{16} singers, we created \num{256} SATB quartets for each piece, which represent all possible combinations of singers taking into account the voice type restriction (i.e. exactly one singer per voice is needed).

The second dataset we use is a proprietary one including \num{26} songs for exactly one singer per part (i.e. \num{4} stems per song), which is a well-suited format for our task as well. All songs offered as part of this dataset are performed in Spanish and their length revolves around two to three minutes, for a total of \num{58} minutes of audio data. 

Our curation work consists in consolidating these two datasets and make sure all the data that we are using remain consistent and well-formatted across all the audio stems, which includes length and amplitude normalization, as well as properties standardization. Most of the files included as part of the initial datasets were presented as \num{10}-seconds long snippets as opposed to full-length songs, which isn't an ideal format to work with. Hence, some additional efforts have been devoted to turn these files in a more convenient and consolidated format.

\section{APPROACH}
\label{sec:method}

Injecting domain knowledge in DNNs has been proved to be an effective way to learn complex input-output relations with high accuracy when the available data happen to be scarce and limited~\cite{INJECT_DK2020}, such as found in our case.

Each of the singing groups in a SATB recordings performs within its own respective frequency range, that is, the voices' F$0$ contour will rarely overlap across the various groups. This factor makes the sources' F$0$ a suitable discriminative feature, which could be injected in the DNN during the training stage and potentially improve the separation of the various singing groups in SATB recordings.

In this view, we propose to adapt the original C-U-Net architecture, which initially embeds the instruments to be separated, \(\overline{z}\), in order to produce the various \textit{FiLM} parameters (\(\gamma, \beta\)), and substitute the external control input data for the F$0$ track of the target source. The new control input vector \(\overline{z}\) will thus hold time as well as frequency dimensions. 

\subsection{Control Input Representation}

As previously mentioned, we use the frame-wise F$0$ as the external control data of our condition generator network. This entails that a few preliminary steps are required prior to proceeding to the training stage. We first automatically extract the sources' F$0$ track using the DIO algorithm~\cite{DIO2009}. Once the raw pitch tracks are obtained, we convert each time-step F$0$ into a one-hot encoded representation as postulated in~\cite{DEEPSR2017}, that is \num{60} frequency bins over \num{6} musical octaves, with a base frequency of \SI{32.7}{\hertz} Hertz, for a total of \num{360} frequency bins per time-step. As a result, for \num{128} time-steps, our control input will be in the shape of \([128,360]\).

\begin{figure}[H]
    \centering
    \includegraphics[width=0.25\textwidth]{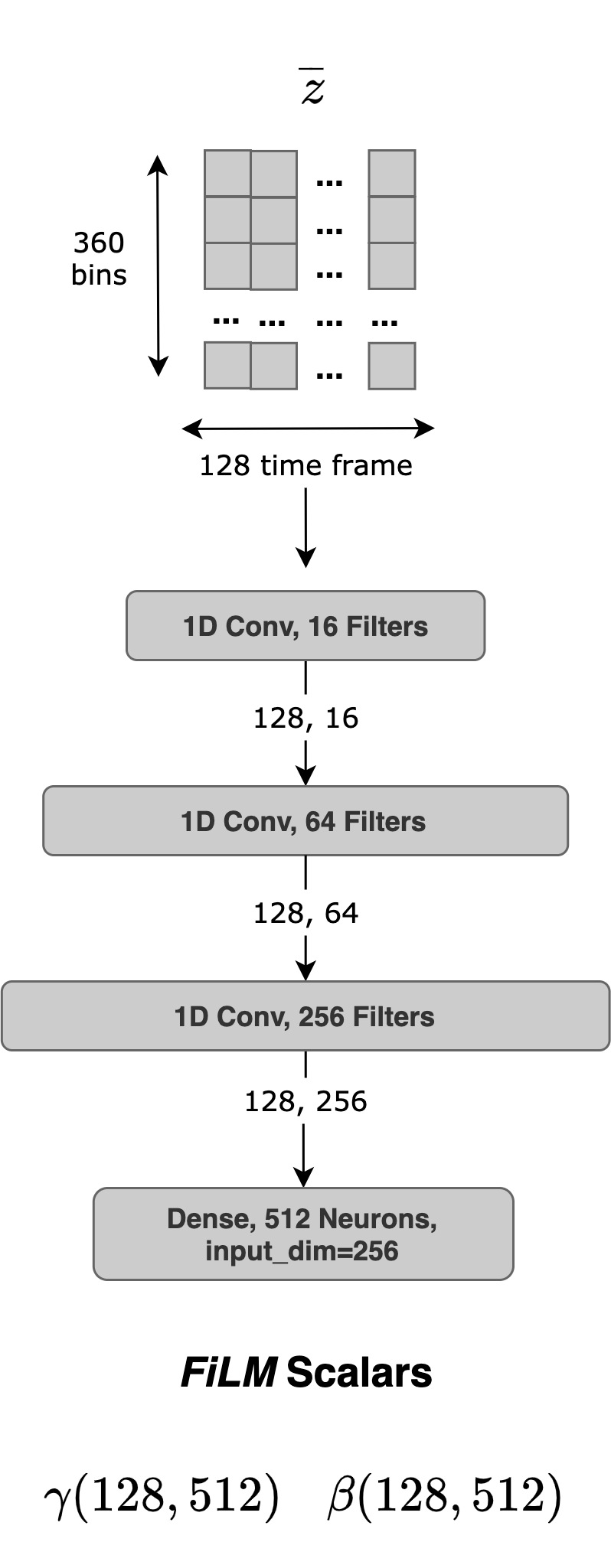}
    \caption{Control model architecture. The convolution is performed across the frequency bins for each time-step. The dense layer provides a specific conditioning for each frequency bin.}
  \label{fig:condition_net}
\end{figure}

\subsection{Control Model} \label{subsec:control_model}

The control model used in our proposed architecture embeds the one-hot encoded CQT F$0$ representation for a given time-step into a set of transforms of identical shape as the spectrogram input. This is achieved by modeling the condition vectors \(\overline{z}\) as \num{1}-D data with multiple feature channels. The condition vectors are then fed into a convolutional neural network (CNN) with a kernel of size \num{10} seizing contextual information from the adjacent time-steps. As a result, all input channels of the initial convolution contribute to all resulting feature maps in the output of the first convolutional layer. Finally a dense layer provides a specific conditioning for each frequency bin at each time-step, taking into account the contextual information previously captured by the CNN. Figure~\ref{fig:condition_net} shows the condition generator architecture in greater details.

\begin{figure}
    \centering
    \includegraphics[width=0.45\textwidth]{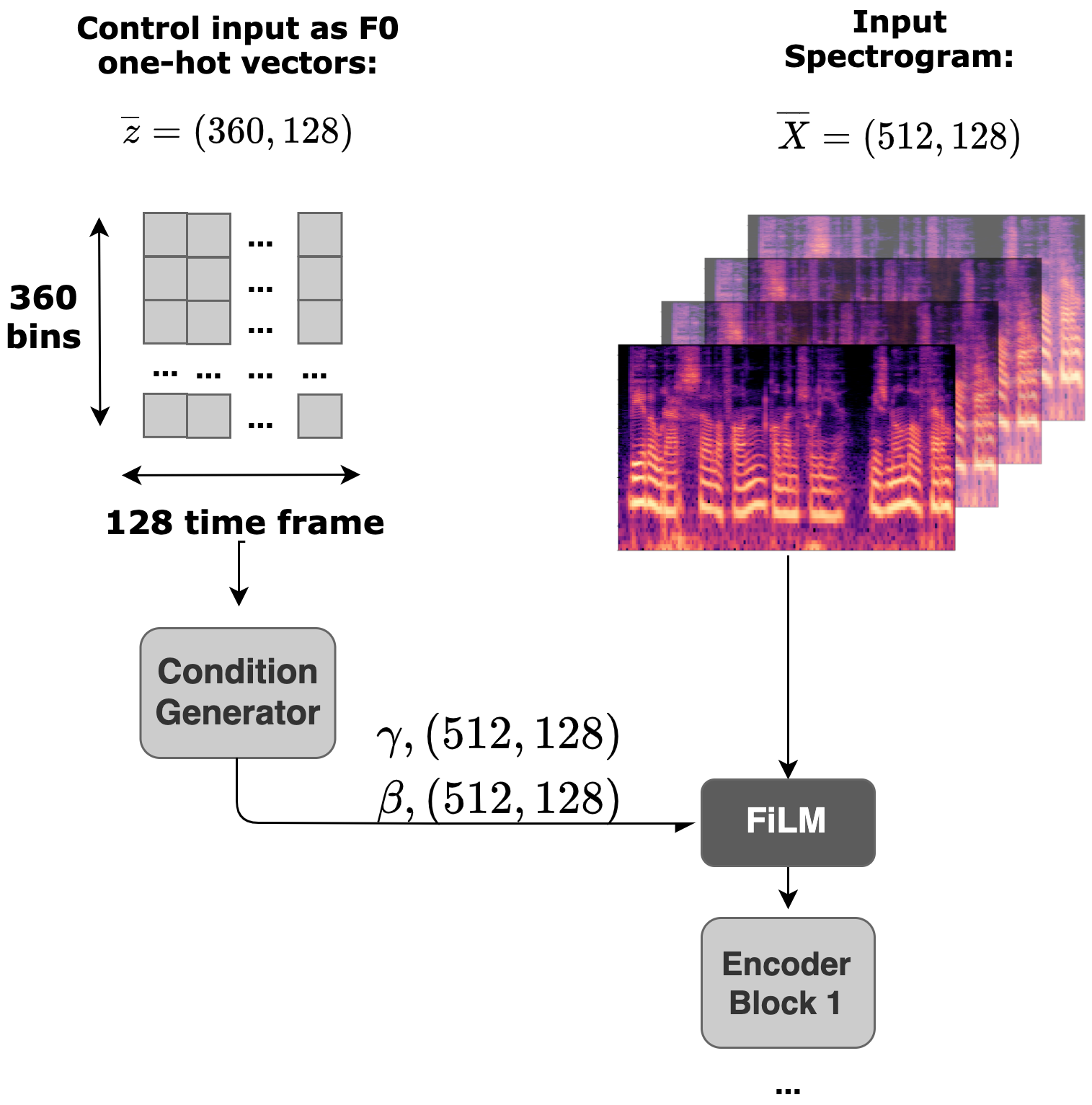}
    \caption{C-U-Net Control Mechanism adapted to our task, with the one-hot vector $\protect\overline{z}$ depicting the various SATB singing groups' F$0$ contour.}
  \label{fig:cunet_mechanism_satb}
\end{figure}

As the temporal relation between the external control input data and the input spectrogram is crucial, it is important to apply these affine transformations while the receptive field of the input is still intact. Hence the \textit{FiLM} layer is applied prior to the encoding path. Figure~\ref{fig:cunet_mechanism_satb} shows the overall structure behind the proposed conditioning architecture.

We propose two variants of the architecture described above, each of which differs slightly in the way it embeds the control input data; the first variant applies a unique affine transform for each individual frequency bin at every input time-step. The resulting scalars in the output of the external CNN model will thus be in the shape \([512,128]\) for a given input spectrogram of the same time context. On the other hand, the second variant applies a single transform for all frequency bins at a given time-step, resulting in a set of scalars of shape \([1,128]\) for 128 spectrogram frames given in the input. We refer to the two approaches as "Domain-Specific Global" and "Domain-Specific Local", respectively and define them with the acronyms \textit{C-U-Net D-S G} and \textit{C-U-Net D-S L} in the rest of this paper. while the three models covered in Section \ref{sec:sota} will be referred to as "Domain-Agnostic" models.

\section{EVALUATION}
\label{sec:eval_methodology}

To assess our proposed approach and show that injecting domain knowledge as control input data to the network improves its performance on SATB recordings, we evaluate the performances of three domain-agnostic state-of-the-art DNN models; U-Net, its waveform adaptation Wave-U-Net, and the original C-U-Net. We then compare the results with the two domain-specific models proposed in Subsection \ref{subsec:control_model}. We evaluate the performance of a model by computing three metrics, SDR, SIR, and SAR~\cite{SiSEC16}, between the predicted and true audio sources. The three measurement metrics describe the overall quality of the separation, the level of interference with other sources as well as the amount of artifact added by the separation algorithm, respectively. The metrics are computed using the \textit{mir\_eval toolbox}~\cite{MIR_EVAL2014} for each of the SATB singing groups. 

\subsection{Train - Test Split}
\label{sec:trainsplit}
Given the limited size of our data, we opted to set apart one song from the proprietary dataset as well as one singer per voice for each song from the CSD in order to build our first use-case test set. The rest of the data was used for training. This allowed us evaluate the model on unseen songs and singers. Our second test case contains unison singing, which was not seen at all during training. As such, we used the three songs from the CSD with all singers for evaluation.

% Because of the limited nature of our dataset, we train these models with mixtures involving exactly one singer per voice (e.g. quartets). To test the model, we use two different test sets; the first set corresponds to four \num{4}-singers songs (quartet), which depicts the type of mixture that the models have been trained on. \darius{Three of those four songs have been extracted from each of the CSD 16-singers songs as quartet songs while the fourth song was taken from our other proprietary dataset. The second set corresponds to \num{16}-singer songs (full choir), representing a mixture setting the models haven't seen. For this set our option remains extremely limited as only the CSD offers this type of mixture setting, we thus opt to include all three CSD songs as part of this second testing set}.

\subsection{Experiment Results}
\label{sec:results}

Subsections~\ref{sec:results_usecase1} and~\ref{sec:results_usecase2} present the performance results for the two test cases described earlier; that is, for the test set involving exactly one singer per part and the other involving exactly four singers per part, respectively. \textit{C-U-Net D-A} refers to the domain-agnostic architecture while \textit{C-U-Net D-S L} and \textit{C-U-Net D-S G} refer to our two domain-specific adaptations. For testing, we use the oracle fundamental frequency of each of the sources, pre-computed prior to model inference. In a complete source separation pipeline, we would complement our system with the multi-pitch algorithm proposed by \cite{CuestaMG20_MultipitchVocalEnsembles_ISMIR}.

\begin{figure*}[t!]
\centering
\begin{subfigure}{1.0\textwidth}
    \centering
    \includegraphics[width=1.0\textwidth]{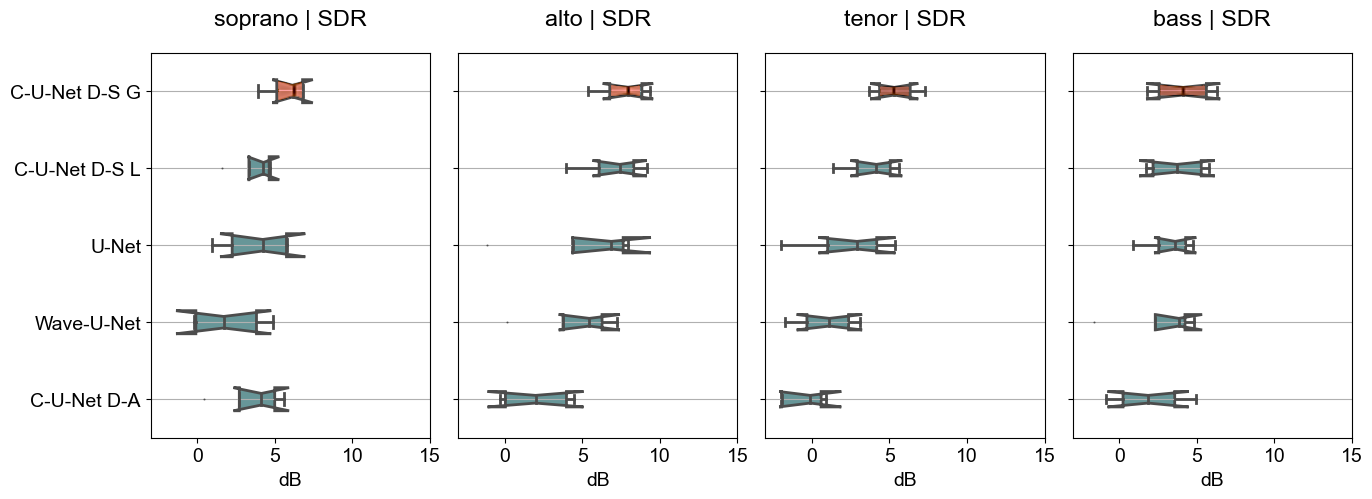}
    \caption{Four-Parts SDR Boxplot Results, Use-Case \num{1}: \num{4}-Singers Mixture}
  \label{fig:sdr_results_case1}
\end{subfigure}
\vspace{0.3cm}
\begin{subfigure}{1.0\textwidth}
    \centering
    \includegraphics[width=1.0\textwidth]{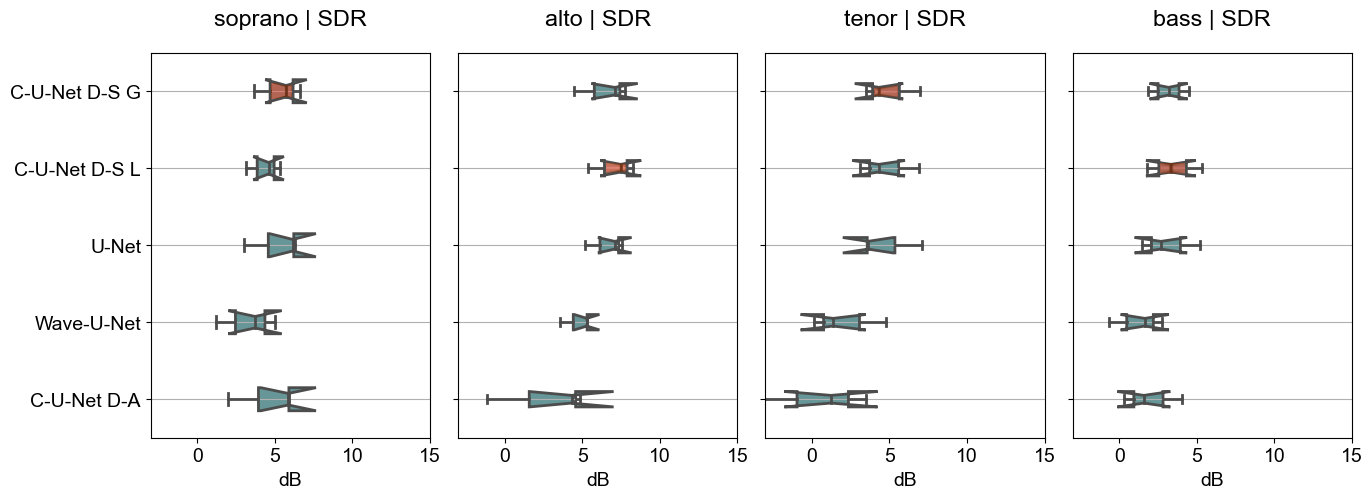}
    \caption{Four-Parts SDR Boxplot Results, Use-Case \num{2}: \num{16}-Singers Mixture}
  \label{fig:sdr_results_case2}
\end{subfigure}
\caption{Boxplot SDR results on the five U-Net based models described in previous sections. Subfigure~\ref{fig:sdr_results_case1} shows the result distribution over the first use-case test set while~\ref{fig:sdr_results_case2} depicts the results for the second use-case. For each one of the SATB parts, the model performing with the highest median is indicated in a dark orange color.}
\label{fig:sdr_results}
\end{figure*}

\begin{table*}[thb!]
\centering
\vspace{0.5cm}
\setlength\tabcolsep{3.0pt}
\begin{tabularx}{\textwidth}{c c c c c c@{\hskip 0.05in} | c c c c c}
    \textbf{Model} & \multicolumn{5}{c}{\textbf{Test Use-Case 1 - SIR (dB)}} 
                   & \multicolumn{5}{c}{\textbf{Test Use-Case 1 - SAR (dB)}} \\
                   
                            & Soprano & Alto & Tenor  & Bass & Avg.
                            & Soprano & Alto & Tenor  & Bass & Avg. \\\cline{2-11}
                            
    \textit{Wave-U-Net}     & 5.99\(\pm\)2.4  & 9.19\(\pm\)2.9  & 4.62\(\pm\)2.1     & 8.49\(\pm\)3.5 & 7.07    & 5.36\(\pm\)1.7 & 7.11\(\pm\)2.4 & 4.79\(\pm\)1.4 & 4.89\(\pm\)1.4 & 5.54 \\
    \textit{U-Net}          & 10.28\(\pm\)2.4 & 10.77\(\pm\)4.1 & 6.70\(\pm\)3.2     & 9.45\(\pm\)2.0 & 9.30    & 5.35\(\pm\)1.8 & 7.13\(\pm\)3.0 & 5.32\(\pm\)1.8 & 4.94\(\pm\)1.1 & 5.69\\
    \textit{C-U-Net D-A}    & 10.09\(\pm\)2.6 & 7.81\(\pm\)1.6  & 3.32\(\pm\)3.3     & 7.61\(\pm\)2.2 & 7.21    & 5.19\(\pm\)1.6 & 4.41\(\pm\)2.8 & 2.65\(\pm\)1.2 & 4.12\(\pm\)2.0 & 4.09\\
    \textit{C-U-Net D-S L}  & 9.71\(\pm\)1.7  & 12.37\(\pm\)1.5 & 9.89\(\pm\)2.2     & 9.71\(\pm\)1.7 & 10.42   & 5.44\(\pm\)1.0 & 8.75\(\pm\)2.0 & 5.58\(\pm\)1.3 & 5.51\(\pm\)1.7 & 6.32\\
    \textit{C-U-Net D-S G}  & \textbf{12.72}\(\pm\)1.8    & \textbf{14.04}\(\pm\)1.5 & \textbf{11.79}\(\pm\)1.5   & \textbf{9.78}\(\pm\)2.1 & \textbf{12.08}    & \textbf{7.02}\(\pm\)1.1    & \textbf{9.02}\(\pm\)1.6 & \textbf{6.86}\(\pm\)1.5   & \textbf{5.93}\(\pm\)1.6 &  \textbf{7.21}\\
    
    \hline
\end{tabularx}
\rule{0pt}{4ex} 
\setlength\tabcolsep{3.0pt}
\begin{tabularx}{1.0\textwidth}{c c c c c c@{\hskip 0.05in} | c c c c c}
            & \multicolumn{5}{c}{\textbf{Test Use-Case 2 - SIR (dB)}} 
            & \multicolumn{5}{c}{\textbf{Test Use-Case 2 - SAR (dB)}}\\
                   
                    & Soprano & Alto & Tenor  & Bass & Avg.
                    & Soprano & Alto & Tenor  & Bass & Avg. \\\cline{2-11}
                    
    \textit{Wave-U-Net}     & 8.13\(\pm\)2.1 & 10.02\(\pm\)0.9 & 6.80\(\pm\)2.2 & 7.45\(\pm\)2.0 & 8.10              & 5.75\(\pm\)1.0 & 6.73\(\pm\)1.1 & 4.79\(\pm\)1.5 & 3.23\(\pm\)0.9 & 5.13\\
    \textit{U-Net}          & \textbf{12.41}\(\pm\)1.8 & 13.11\(\pm\)1.2 & 10.26\(\pm\)1.1 & 8.50\(\pm\)2.3 & 11.07    & 6.31\(\pm\)1.4 & 7.97\(\pm\)1.0 & \textbf{6.59}\(\pm\)1.8 & 5.27\(\pm\)1.1 & 6.54\\
    \textit{C-U-Net D-A}    & 11.99\(\pm\)1.8 & 9.08\(\pm\)2.8 & 5.65\(\pm\)3.1 & 7.60\(\pm\)2.0 & 8.58                & 5.77\(\pm\)1.7 & 4.60\(\pm\)2.9 & 3.39\(\pm\)1.8 & 4.15\(\pm\)1.2 & 4.48\\
    \textit{C-U-Net D-S L}  & 10.32\(\pm\)1.1 & 13.06\(\pm\)1.7 & 10.77\(\pm\)1.5 & 8.89\(\pm\)2.2 & 10.76             & 6.02\(\pm\)0.9 & \textbf{8.59}\(\pm\)1.2 & 6.45\(\pm\)1.7 & \textbf{5.59}\(\pm\)1.1 & \textbf{6.66}\\
    \textit{C-U-Net D-S G}  & 12.08\(\pm\)1.5 & \textbf{13.50}\(\pm\)2.5 & \textbf{12.05}\(\pm\)1.3 & \textbf{8.91}\(\pm\)2.1 & \textbf{11.63}    & \textbf{6.68}\(\pm\)1.2 & 7.70\(\pm\)1.3 & 6.17\(\pm\)1.6 & 5.17\(\pm\)0.8 & 6.43\\
    
    \hline
\end{tabularx}

\caption{SIR and SAR mean and standard deviation results on the four SATB parts as well as their average for the five U-Net based models described in previous sections. The top table depicts the results obtained from the first use-case test set while the bottom one the second use-case test set.}
\captionsetup{width=1.0\linewidth }
\label{table:results}
\end{table*}

\subsubsection{Use-Case 1:}
\label{sec:results_usecase1}

Table~\ref{table:results} portrays the mean SIR and SAR results on all the SATB parts and average for all five models mentioned in previous sections. Figure~\ref{fig:sdr_results_case1} details the SDR score distributions on our first test set. We observe that our two adaptations, \textit{C-U-Net D-S L} and \textit{C-U-Net D-S G}, call attention to a significant score gap between domain-agnostic and domain-specific models, with an average increase of about \num{1}dB SDR and \num{1.5}dB SIR between the two different approaches. These improvements underline an overall better quality of the predicted sources (SDR) as well as a decline in interference between the various predictions. Our domain-specific architecture hence demonstrates a better ability to cope with the correlated nature of the various SATB sources and seem to predict an appropriate spectral mask for each of them. We also observe that our proposed adaptations return the lowest SDR, SIR and SAR performances for the Bass part, specifically. This could be due to the fact that the Bass group, among all SATB groups, shares the highest number of harmonics with its other source counter parts.

We also note that the mean SDR substantially drops for the Tenor singing group across nearly all domain-agnostic models, reaching a negative result with the \textit{C-U-Net D-A} architecture (\num{-1.25}dB SDR). We speculate that the reason behind such decline can be directly related to the close nature of the F$0$ contours of both Alto and Tenor singing groups, making it harder for domain-agnostic architecture to distinguish between the two sources. This limitation brings yet another justification for the conditioning approach we have taken in this paper. 

\subsubsection{Use-Case 2:}
\label{sec:results_usecase2}

Figure \ref{fig:sdr_results_case2} as well as the bottom portion of Table~\ref{table:results} presents the SDR, SIR and SAR scores on the second use-case test set. We observe that the introduction of more complex mixtures involving a higher number of singers (i.e.: \num{16} singers in this case) decreases the performance of our proposed models, with an average SDR barely surpassing U-Net's for our \textit{C-U-Net D-S G} model and levelling it out for the \textit{C-U-Net D-S L} model. This can be attributed to the use of  the mean of the various pitches present in a singing group, to represent the pitch of the unison. Since domain-agnostic models, such as the plain U-Net, don't hold this assumption, these architectures are less prone to errors when exposed to these type of mixture settings~\footnote{Audio examples showcasing how the inference of our proposed models compare against other state-of-the-art architectures are available online. On the same page is also included a link to our pre-trained models: \url{https://darius522.github.io/satb-source-separation-results/}}.

\section{CONCLUSIONS AND FUTURE WORK}
\label{sec:conclusion}

In this work we have presented the task of musical source separation applied to SATB choir recordings. We first described the consolidated dataset that we've specifically curated for this experiment and its potential use and application for future related research. We then assessed how well recent domain-agnostic deep learning based architectures for musical source separation performed on this task, given two different use-cases; \num{4}-singers mixture and \num{16}-singers mixture separation. An adaptation of the U-Net architecture was then proposed, consisting in conditioning some of the network parameters on the fundamental frequency contour of each of the SATB mixture sources. The preliminary results showed that taking advantage of domain-knowledge during the training process improved the performance on both of our proposed use-cases.
For the evaluation presented in this paper, we use the oracle F$0$ is currently used as external control input data to the network. In a complete source separation pipeline, we plan on combining the task of multi-pitch tracking~\cite{CuestaMG20_MultipitchVocalEnsembles_ISMIR} with the system presented in this paper. We also plan on validating our evaluation with perceptual listening tests and exploring applications of the SATB separation. including remixing,  transcription and transposition combined with the work presented in \cite{Chandna20_ISMIR, chandna2020content}. 

\section{acknowledgements}
The TITANX used for this research was donated by the NVIDIA Corporation. This work is partially supported by the Towards Richer Online Music Public-domain Archives (TROMPA H2020 770376) project. Helena Cuesta is supported by the FI Predoctoral Grant from AGAUR (Generalitat de Catalunya). The authors would like to thank Rodrigo Schramm and Emmanouil Benetos for sharing their singing voice datasets for this research.

% For bibtex users:
\bibliography{ISMIRtemplate}

\begin{thebibliography}{10}
\providecommand{\url}[1]{#1}
\csname url@samestyle\endcsname
\providecommand{\newblock}{\relax}
\providecommand{\bibinfo}[2]{#2}
\providecommand{\BIBentrySTDinterwordspacing}{\spaceskip=0pt\relax}
\providecommand{\BIBentryALTinterwordstretchfactor}{4}
\providecommand{\BIBentryALTinterwordspacing}{\spaceskip=\fontdimen2\font plus
\BIBentryALTinterwordstretchfactor\fontdimen3\font minus
  \fontdimen4\font\relax}
\providecommand{\BIBforeignlanguage}[2]{{%
\expandafter\ifx\csname l@#1\endcsname\relax
\typeout{** WARNING: IEEEtran.bst: No hyphenation pattern has been}%
\typeout{** loaded for the language `#1'. Using the pattern for}%
\typeout{** the default language instead.}%
\else
\language=\csname l@#1\endcsname
\fi
#2}}
\providecommand{\BIBdecl}{\relax}
\BIBdecl

\bibitem{SATB_RANGE}
M.~Scirea and J.~A. Brown, ``{Evolving four part harmony using a multiple
  worlds model},'' in \emph{Proceedings of the 7th International Joint
  Conference on Computational Intelligence (IJCCI)}, vol.~1.\hskip 1em plus
  0.5em minus 0.4em\relax IEEE, 2015, pp. 220--227.

\bibitem{MONO_SIGSEP2017}
P.~Chandna, M.~Miron, J.~Janer, and E.~G\'omez", ``{Monoaural Audio Source
  Separation Using Deep Convolutional Neural Networks},'' in
  \emph{International Conference on Latent Variable Analysis and Signal
  Separation}, 2017, pp. 258--266.

\bibitem{UMIX2019}
\BIBentryALTinterwordspacing
F.-R. St{\"o}ter, S.~Uhlich, A.~Liutkus, and Y.~Mitsufuji, ``Open-unmix - a
  reference implementation for music source separation,'' \emph{Journal of Open
  Source Software}, 2019. [Online]. Available:
  \url{https://doi.org/10.21105/joss.01667}
\BIBentrySTDinterwordspacing

\bibitem{ENTOEND2018}
\BIBentryALTinterwordspacing
F.~Lluís, J.~Pons, and X.~Serra, ``{End-to-End Music Source Separation: Is it
  Possible in the Waveform Domain?}'' in \emph{Proceedings of Interspeech
  2019}, September 2019. [Online]. Available:
  \url{http://dx.doi.org/10.21437/interspeech.2019-1177}
\BIBentrySTDinterwordspacing

\bibitem{SiSEC16}
A.~Liutkus, F.-R. St{\"o}ter, Z.~Rafii, D.~Kitamura, B.~Rivet, N.~Ito, N.~Ono,
  and J.~Fontecave, ``The 2016 signal separation evaluation campaign,'' in
  \emph{International Conference on Latent Variable Analysis and Signal
  Separation}.\hskip 1em plus 0.5em minus 0.4em\relax Springer, 2017, pp.
  323--332.

\bibitem{MUSDB18}
\BIBentryALTinterwordspacing
Z.~Rafii, A.~Liutkus, F.-R. St{\"o}ter, S.~I. Mimilakis, and R.~Bittner, ``The
  {MUSDB18} corpus for music separation,'' December 2017. [Online]. Available:
  \url{https://doi.org/10.5281/zenodo.1117372}
\BIBentrySTDinterwordspacing

\bibitem{UNET2015}
O.~Ronneberger, P.~Fischer, and T.~Brox, ``U-net: Convolutional networks for
  biomedical image segmentation,'' in \emph{Proceedings of the International
  Conference on Medical image computing and Computer-assisted
  Intervention}.\hskip 1em plus 0.5em minus 0.4em\relax Springer, 2015, pp.
  234--241.

\bibitem{WAVEUNET2018}
D.~Stoller, S.~Ewert, and S.~Dixon, ``{Wave-U-Net: A Multi-Scale Neural Network
  for End-to-End Audio Source Separation},'' in \emph{Proceedings of the
  International Society for Music Information Retrieval Conference (ISMIR)},
  2018.

\bibitem{CUNET2019}
G.~Meseguer-Brocal and G.~Peeters, ``Conditioned u-net: Introducing a control
  mechanism in the u-net for multiple source separations.'' November 2019.

\bibitem{SING_UNET2017}
A.~Jansson, E.~Humphrey, N.~Montecchio, R.~Bittner, A.~Kumar, and T.~Weyde,
  ``{Singing Voice Separation with Deep U-Net Convolutional Networks},'' in
  \emph{Proceedings of the International Society for Music Information
  Retrieval Conference (ISMIR)}, 2017.

\bibitem{FILM2017}
E.~Perez, F.~Strub, H.~de~Vries, V.~Dumoulin, and A.~C. Courville, ``{FiLM:
  Visual Reasoning with a General Conditioning Layer},'' in \emph{Proceedings
  of the 32nd AAAI Conference on Artificial Intelligence}, April 2018.

\bibitem{CUESTA_DST2018}
H.~Cuesta, E.~G{\'o}mez, A.~Martorell, and F.~Lo{\'a}iciga, ``{Analysis of
  Intonation in Unison Choir Singing},'' in \emph{Proceedings of the
  International Conference of Music Perception and Cognition ({ICMPC})}, Graz
  (Austria), July 2018, pp. 125--130.

\bibitem{MULTIF0_SATB2019}
H.~Cuesta, E.~G\'omez, and P.~Chandna, ``{A Framework for Multi-f0 Modeling in
  SATB Choir Recordings},'' in \emph{{Proceedings of the Sound and Music
  Computing (SMC) Conference}}, M\'alaga (Spain), April 2019.

\bibitem{INJECT_DK2020}
M.~Silvestri, M.~Lombardi, and M.~Milano, ``Injecting domain knowledge in
  neural networks: a controlled experiment on a constrained problem,''
  \emph{ArXiv}, vol. abs/2002.10742, 2020.

\bibitem{DIO2009}
\BIBentryALTinterwordspacing
M.~Morise, H.~Kawahara, and H.~Katayose, ``Fast and reliable f0 estimation
  method based on the period extraction of vocal fold vibration of singing
  voice and speech,'' in \emph{AES 35th International Conference on Audio for
  Games}, February 2009. [Online]. Available:
  \url{http://www.aes.org/e-lib/browse.cfm?elib=15165}
\BIBentrySTDinterwordspacing

\bibitem{DEEPSR2017}
R.~M. Bittner, B.~McFee, J.~Salamon, P.~Li, and J.~P. Bello, ``{Deep Salience
  Representations for F0 Estimation in Polyphonic Music},'' in
  \emph{Proceedings of the International Society for Music Information
  Retrieval Conference, (ISMIR)}, 2017.

\bibitem{MIR_EVAL2014}
C.~Raffel, B.~Mcfee, E.~J. Humphrey, O.~N. Justin~Salamon, D.~Liang, and
  D.~P.~W. Ellis, ``mir\_eval: a transparent implementation of common mir
  metrics,'' in \emph{Proceedings of the International Society for Music
  Information Retrieval Conference (ISMIR)}, 2014.

\bibitem{CuestaMG20_MultipitchVocalEnsembles_ISMIR}
H.~Cuesta, B.~McFee, and E.~G\'omez, ``{Multiple F0 Estimation in Vocal
  Ensembles using Convolutional Neural Networks},'' in \emph{Proceedings of the
  21st International Society for Music Information Retrieval Conference
  (ISMIR)}, Montreal, Canada (Virtual), 2020.

\bibitem{Chandna20_ISMIR}
P.~Chandna, H.~Cuesta, and E.~G\'omez, ``{A Deep Learning Based
  Analysis-Synthesis Framework For Unison Singing},'' in \emph{Proceedings of
  the 21st International Society for Music Information Retrieval Conference
  (ISMIR)}, Montreal, Canada (Virtual), 2020.

\bibitem{chandna2020content}
P.~Chandna, M.~Blaauw, J.~Bonada, and E.~G{\'o}mez, ``Content based singing
  voice extraction from a musical mixture,'' in \emph{{Proceedings of the 45th
  IEEE International Conference on Acoustics, Speech, and Signal Processing
  (ICASSP)}}.\hskip 1em plus 0.5em minus 0.4em\relax IEEE, 2020, pp. 781--785.

\end{thebibliography}

% For non bibtex users:
%\begin{thebibliography}{citations}
% \bibitem{Author:17}
% E.~Author and B.~Authour, ``The title of the conference paper,'' in {\em Proc.
% of the Int. Society for Music Information Retrieval Conf.}, (Suzhou, China),
% pp.~111--117, 2017.
%
% \bibitem{Someone:10}
% A.~Someone, B.~Someone, and C.~Someone, ``The title of the journal paper,''
%  {\em Journal of New Music Research}, vol.~A, pp.~111--222, September 2010.
%
% \bibitem{Person:20}
% O.~Person, {\em Title of the Book}.
% \newblock Montr\'{e}al, Canada: McGill-Queen's University Press, 2020.
%
% \bibitem{Person:09}
% F.~Person and S.~Person, ``Title of a chapter this book,'' in {\em A Book
% Containing Delightful Chapters} (A.~G. Editor, ed.), pp.~58--102, Tokyo,
% Japan: The Publisher, 2009.
%
%
%\end{thebibliography}

\end{document}